\documentclass[twocolumn,superscriptaddress, aps,pra ]{revtex4-1}
\makeatletter
\newcommand*{\rom}[1]{\expandafter\@slowromancap\romannumeral #1@}
\makeatother
\usepackage{color}
\usepackage{graphicx}
\usepackage{hyperref}
\usepackage{amsmath,amssymb}
\usepackage{epstopdf}
\usepackage{subcaption}
\graphicspath{{figs/}}
\begin{document}
\title{  Effects of high-intensity Lasers on the Entanglement fidelity of quantum plasmas}

\author{R. Roozehdar Mogaddam}
\affiliation{Department of Physics, Ferdowsi University of Mashhad,   P.O. Box 1436 Mashhad, Iran}
\author {N. Sepehri Javan}
\email{sepehri${_}$javan@uma.ac.ir}
\affiliation{Department of Physics, University of Mohaghegh Ardabili, P.O. Box 179, Ardabil, Iran}

\author{K. Javidan}
\affiliation{Department of Physics, Ferdowsi University of Mashhad,   P.O. Box 1436 Mashhad, Iran}

\author{H. Mohammadzadeh}
\affiliation{Department of Physics, University of Mohaghegh Ardabili, P.O. Box 179, Ardabil, Iran}

\date{2019/11/01}

\begin{abstract}
The dynamics of entanglement during the low energy scattering processes in bi-partite systems at the presence of a laser field is studied, using the Kramers-Henneberger unitary transformation as the semiclassical counterpart of the Block-Nordsieck transformation, in the quantizied field formalism. The Stationary-state Schrodinger equation for quantum scattering process is obtained for such systems. Then, by using partial wave analysis, we introduce new form of entanglement fidelity containing high-intense laser field. Therefore, the effective potential of hot quantum plasmas including plasmon and quantum screening effect is used to show entanglement fidelity ratio as a function of the laser amplitude, plasmon parameter and the Debye length parameter for elastic electron-ion collisions. It is shown that the amplitude of laser beam or free electron oscillation play important roles in the evolution of entanglement of the system. 

\end{abstract}


\maketitle

\section{Introduction}\label{1}
The quantum correlation among distinct quantum systems, (or entanglement) is a complex and powerful phenomenon in the framework of quantum mechanics which nowadays is widely used in all branches of science \cite{xiang2007entanglement,hamma2008entanglement,prants2006entanglement}. Produced entanglement due to the Lorentz invariance violation in high energy physics processes \cite{bernabeu2011t}, expansion of universe \cite{ball2006entanglement, fuentes2010entanglement, friis2013entanglement, mohammadzadeh2015entanglement, farahmand2017quantum, mohammadzadeh2017entropy}, variation of entanglement due to the environmental interactions \cite{zyczkowski2001dynamics, nha2004entanglement, schliemann2003electron} effects of observer acceleration in entanglement degradation (the Unruh effect) \cite{fuentes2005alice, downes2011entangling, mehri2011pseudo, farahmand2017residual}, the Quantum self-organization applied in physics \cite{otsuka2018quantum} and biophysics \cite{rakovic2007macroscopic} are some examples in this research area.

However, there have been presented many works in theoretical features of entanglement \cite{wei2003geometric,yang2009squashed}, but there exist few experimental based researches in this area \cite{walborn2006experimental,lee2019nonlocal}, because of many difficulties in measuring the quantum entanglement in laboratory and needing very high technological tools. One of successful experimental methods is measuring the entanglement in bi-particle interaction.  Saraga et. al. \cite{saraga2004coulomb} observed creation of  EPR pairs \cite{einstein1935quantum} during the 2-dimensional Coulomb scattering in an electron gas. They found that created entanglement is very sensitive to the potential energy function of bi-electron system. Measuring of entanglement in terms of wave packet localization has been investigated by Fedorov et al \cite{fedorov2004packet}, while Mishima and colleagues \cite{mishima2004entanglement} have found a suitable measure for entanglement during the scattering interactions which is now called as Fidelity Entanglement (EF). The EF is now widely used because of its effectiveness in realization of quantum entanglement and quantifying information processes \cite{hong2014electron,falaye2019entanglement}. 
     
Several inter-particle interactions which are described by different functions of potential energies, produce different degrees of entanglement. By considering such descriptions, we can clearly understand that plasma is a very ideal media for creation and measuring some features of entanglement. It is because of the presence of high density charged particles. It is clear that in such media collective effects via long-range electromagnetic interactions play an important role and thus, quantum correlations can be appeared \cite{mogaddam2019entanglement}. Presenting a successful definition of an appropriate potential which includes all important features on interaction is the first and also most important step in studying the quantum entanglement in plasmas. There are several theoretical and experimental investigations which provide more accurate knowledge about the nature of potential energy function in classical, semi-classical and quantum plasmas. 
   
The simplest description for interaction of charged particles in plasma is presented using the Debye-Huckel screened potential which is applied in ideal plasmas where the energy if inter-particle interactions is smaller than the average kinetic energy of plasma constituents \cite{kvasnica1973bremsstrahlung,ramazanov2001coulomb}.

The effective interaction potential between projectile electrons and dressed ions by considering the strong quantum recoil effects in quantum plasmas has been investigated in \cite{khan2015wake}. Shukla and Eliasson \cite{shukla2012novel} has presented a practical definition for the interaction potential function in degenerate electron-ion quantum plasmas by considering the screening and electron exchange effects. The effective potential in electron-ion interaction by considering the standard Debye potential as well as the effective Friedel far-field interaction term between particles for collisional degenerate quantum plasmas have been presented in \cite{smirnov2008plasma}. An acceptable form for interaction potential in strongly coupled semi-classical plasmas can be constructed by effective pseudo-potential \cite{ramazanov2002effective} interaction between different kinds of plasma particles based on the dielectric response function analysis. The inter-particle interaction potential for dressed electron-ion scattering in hot quantum plasmas \cite{ramazanov2005runaway} as well as in dense quantum plasmas with different initial conditions has been investigated \cite{ramazanov2015effective,reinholz1995thermodynamic,vorberger2013equation}. Quantum mechanical effects based on the definition of effective potential functions are widely investigated in dense plasmas appeared in the core of giant planets \cite{french2012ab}, high energy laser-solid plasma interaction \cite{,marklund2006nonlinear,kremp1999quantum,lower1994uniform,roozehdar2018perturbative,roozehdar2019modulation}, ion accelerator \cite{tahir2011generation}, compact astrophysical objects and cosmological environments \cite{opher2001nuclear,jung2001quantum}, construction of ultra-small electronic devices \cite{markowich1990semiconductor}, metal nanostructures \cite{manfredi2005model} and many other subjects.

The entanglement fidelity for various types of interaction potentials have been calculated \cite{jung2011plasmon,chang2005collective,lee2017collisional,shin2008nonthermal}. Falaye and colleagues \cite{falaye2019entanglement} have investigated the FE at the presence of external magnetic field. In this work, we have extended this problem by including effects of electric potential produced by applied laser field in considered plasma media. It is clear that, without considering the electric field due to the applied laser, results cannot be described correctly. Thus, we expect to fine more accurate results by adding this issue. Indeed, the laser field effectively change the interaction potential in the system And thus we expect to find some important changes in dynamics of entanglement in the system which can be described by the FE. 

The paper is organized as follows: We summarize the bi-partite scattering process and obtained Stationary-state Schrodinger equation by considering laser effect in Sec. \ref{2}. EF at the presence of a laser field represent in Sec. \ref{3}. Effective potential for a hot quantum plasma is introduced in Sec. \ref{4}. we focus on deriving analytical relationships and evaluate the EF for hot quantum plasmas in Sec.  \ref{5}. In section \ref{6}, we have obtained some Numerical discussion about influence of laser effect and other parameter on EF. Finally, we conclude the paper in Sec. \ref{7}.

	\section{Theory and Calculations}\label{2}
We consider the bi-partite scattering of particles in the framework of non-relativistic quantum mechanics. Total Hamiltonian of the system is constructed for two-particles in the center of mass system in the presence of linearly polarized intense laser field radiation as follows \cite{sakurai1995modern}:  
	
	\begin{align}\label{eq1}
		\hat H = \frac{1}{{2\mu }}{\left( {p + e{\bf{A}}\left( {{\bf{r}},t} \right)} \right)^2} - e\phi \left( {{\bf{r}},t} \right) + V\left( {\bf{r}} \right),
	\end{align}	
where $\phi \left( {{\bf{r}},t} \right)$ and ${\bf{A}}\left( {{\bf{r}},t} \right)$ are scalar and vector potential of the laser beam respectively, which are invariant under the gauge transformation. $\mu $ and $V\left( {\bf{r}} \right)$ are the effective mass and the potential energy between the particles. Also, we consider spherically symmetric potential energy in our work. 

In order to derive the equation of motion for two particles interacted by a spherically symmetric potential under the influence of linearly polarized intense laser, we have to solve time-dependent Schrodinger wave equation: $i\hbar \partial \psi \left( {{\bf{r}},t} \right)/\partial t = \hat H\psi \left( {{\bf{r}},t} \right)$ where Hamiltonian is given by the Eq. (\ref{eq1}). Thus, we have:

	\begin{align}\label{eq2}
	i\hbar \frac{\partial }{{\partial t}}\psi &\left( {{\bf{r}},t} \right) = \left[ { - \frac{{{\hbar ^2}}}{{2\mu }}{\nabla ^2} - i\hbar \frac{e}{{2\mu }}} \right.\left( {{\bf{A}}\left( {{\bf{r}},t} \right).\nabla  + \nabla .{\bf{A}}\left( {{\bf{r}},t} \right)} \right)\nonumber\\
	&\left. {+ \frac{{{e^2}}}{{2\mu }}{\bf{A}}{{\left( {{\bf{r}},t} \right)}^2} - e\phi \left( {{\bf{r}},t} \right) + V\left( {\bf{r}} \right)} \right]\psi \left( {{\bf{r}},t} \right),
	\end{align}	

we consider the Coulomb gauge \cite{rousseaux2005lorenz}, such that $\nabla .{\bf{A}}\left( {{\bf{r}},t} \right) = 0$ (and $\phi \left( {{\bf{r}},t} \right) = 0$) in empty space and simplify the interaction term in Eq. \ref{eq2} by applying gauge transformations within the framework of dipole approximation. In this approximation, for an atom which is located at the position ${r_0}$, the vector potential can be considered independent of space, such that ${\bf{A}}\left( {{\bf{r}},t} \right) \approx {\bf{A}}\left( t \right)$. Moreover, term ${\bf{A}}{\left( {{\bf{r}},t} \right)^2}$ appearing in equation (\ref{eq2}) is noticeable only in media with extremely high field strength and thus this term usually is very small and can be eliminated by extracting a time dependent phase factor from the wave function via \cite{burnett1993atoms}
	\begin{align}\label{eq3}
\varphi \left( {{\bf{r}},t} \right) = \exp \left[ {\frac{{i{e^2}}}{{2\mu \hbar }}\int_{ - \infty }^t {{\bf{A}}{{\left( {t'} \right)}^2}dt'} } \right]\psi \left( {{\bf{r}},t} \right),
\end{align}	
Now we need to calculate the velocity gauge. According to equation (\ref{eq3}), we can write: 
	\begin{align}\label{eq4}
i\hbar \frac{\partial }{{\partial t}}\varphi \left( {{\bf{r}},t} \right) = \left[ { - \frac{{{\hbar ^2}}}{{2\mu }}{\nabla ^2} - i\hbar \frac{e}{\mu }{\bf{A}}\left( t \right).\nabla  + V\left( {\bf{r}} \right)} \right]\varphi \left( {{\bf{r}},t} \right),
\end{align}	
But, in order to find the state function of a system composed of two particles under intense high-frequency laser field we have to transform equation (\ref{eq4}) into the Kramers–Henneberger (K–H) accelerated frame \cite{henneberger1968perturbation,wei2017pursuit}. The K–H frame is a reference frame in which a free electron moves at the influence of an applied laser field \cite{reed1990ionization}. The wave functions in the laboratory and K–H frames are related by a unitary transformation as \cite{reed1990ionization}: 
	\begin{align}\label{eq5}
\Psi \left( {{\bf{r}},t} \right) = {U^\dag }\varphi \left( {{\bf{r}},t} \right),
\end{align}	
where $U = \exp \left( { - \frac{i}{\hbar }{\bf{\xi }}\left( t \right).{\bf{p}}} \right)$ is local unitary transformation of the particle where ${\bf{\xi }}\left( t \right) = \frac{e}{\mu }\int_{}^t {{\bf{A}}\left( {t'} \right)} \,dt'$ represents a shift to the accelerated frame of reference. It is indeed semi-classical counterpart of the Block–Nordsieck transformation in the quantized field formalism, so that the coupling term $ {\bf{A}}\left( t \right).\nabla $ in the velocity gauge (i.e. equation \ref{eq4})) is eliminated. More explicitly, this can be done via
	\begin{align}\label{eq6}
i\hbar {U^\dag }\frac{\partial }{{\partial t}}U\Psi \left( {{\bf{r}},t} \right) &= {U^\dag }\left[ { - \frac{{{\hbar ^2}}}{{2\mu }}{\nabla ^2}} \right.\nonumber\\
&\left. { - i\hbar \frac{e}{\mu }{\bf{A}}\left( t \right).\nabla  + V\left( {\bf{r}} \right)} \right]U\Psi \left( {{\bf{r}},t} \right),
\end{align}	
Evaluation of terms in equation \ref{eq6} are straightforward. The term ${U^\dag }V\left( {\bf{r}} \right)U$ can be evaluated as following:
	\begin{align}\label{eq7}
{U^\dag }V&\left( {\bf{r}} \right)U = \exp \left( {\frac{i}{\hbar }{\bf{\xi }}\left( t \right).{\bf{p}}} \right)V\left( {\bf{r}} \right)\exp \left( { - \frac{i}{\hbar }{\bf{\xi }}\left( t \right).{\bf{p}}} \right)\nonumber\\ 
  &=V\left( {\bf{r}} \right) + \left[ {{\bf{\xi }}\left( t \right).\nabla } \right]V\left( {\bf{r}} \right) + \frac{1}{{2!}}{\left[ {{\bf{\xi }}\left( t \right).\nabla } \right]^2}V\left( {\bf{r}} \right) + ...\nonumber\\
  &=V\left( {{\bf{r}} + {\bf{\xi }}\left( t \right)} \right)
\end{align}	
where ${\bf{\xi }}\left( t \right)$ represents the displacement of a free electron under the influence of the incident laser field. Hence, equation \ref{eq6} becomes
	\begin{align}\label{eq8}
i\hbar \frac{\partial }{{\partial t}}\Psi \left( {{\bf{r}},t} \right) = \left[ { - \frac{{{\hbar ^2}}}{{2\mu }}{\nabla ^2} + V\left( {{\bf{r}} + {\bf{\xi }}\left( t \right)} \right)} \right]\Psi \left( {{\bf{r}},t} \right)
\end{align}
Equation (\ref{eq8}) is a space-translated version of the time-dependent Schrödinger wave equation with incorporation of ${\bf{\xi }}\left( t \right)$ into the potential in order to simulate the interaction of atomic system with the laser field. Now, by considering the steady field condition, the vector potential takes the form $A\left( t \right) = {E_0}{\omega ^{ - 1}}\cos \left( {\omega t} \right)$ with $\xi \left( t \right) = {\xi _0}\sin \left( {\omega t} \right)$, where ${\xi _0} = e{E_0}/\mu {\omega ^2}$ is the amplitude of oscillation of a free electron in the field (which is called the laser-dressing parameter),${E_0}$ denotes the amplitude of electromagnetic field strength and $\omega $ is the angular frequency. Now, we consider a pulse of electric field with a steady amplitude, the wave function in the K-H frame represents by the following Floquet form \cite{burnett1993atoms}:
	\begin{align}\label{eq9}
\Psi \left( {{\bf{r}},t} \right) = {e^{ - \frac{{i{E_{KH}}}}{\hbar }t}}\sum\limits_n {\Psi _k^n\left( {\bf{r}} \right)} \,{e^{ - in\omega t}},
\end{align}
where Floquet quasi-energy has been denoted by ${E_{KH}}$. The potential in the K-H frame can be expanded in Fourier series as \cite{yesilgul2012effects,kasapoglu2008effects,fonseca1994intense}: 
	\begin{align}\label{eq10}
V\left( {{\bf{r}} + {\bf{\xi }}\left( t \right)} \right) = \sum\limits_{m =  - \infty }^{ + \infty } {{V_m}\left( {{\xi _0},r} \right)} \,{e^{ - im\omega t}},
\end{align}
	\begin{align}\label{eq11}
{V_m}\left( {{\xi _0},r} \right) = \frac{{{i^m}}}{\pi }\int_{ - 1}^{ + 1} {{V_m}\left( {r + {\xi _0}\rho } \right)} \frac{{{T_n}\left( \rho  \right)}}{{\sqrt {1 - {\rho ^2}} }}d\rho ,
\end{align}
where we have taken the period as $2\pi /\omega $ and introduced a new transformation of the form $\rho  = \sin \left( {\omega \,t} \right)$ while ${T_n}\left( \rho  \right)$ are Chebyshev polynomials. Substituting equations (\ref{eq9}), (\ref{eq10}) and (\ref{eq11}) into (\ref{eq8}) yields a set of coupled differential equations:
	\begin{align}\label{eq12}
\left[ { - \frac{{{\hbar ^2}}}{{2\mu }}{\nabla ^2} + {V_m}\left( {{\xi _0},r} \right) - \left( {{E_{KH}} + n\hbar \omega } \right)} \right]\Psi _k^m\left( {\bf{r}} \right) =\nonumber\\  - \sum\limits_{m =  - \infty }^{ + \infty } {{V_{n - m}}} \Psi _k^m\left( {\bf{r}} \right),
\end{align}
Considering the lowest order of approximation $\left( {n = 0} \right)$ and high frequency limit condition (which means ${V_m}$ vanishes when $m \ne 0$), equation \ref{eq12} becomes:
	\begin{align}\label{eq13}
\left[ { - \frac{{{\hbar ^2}}}{{2\mu }}{\nabla ^2} + {V_0}\left( {{\xi _0},r} \right) - {E_{KH}}} \right]\Psi _k^0\left( {\bf{r}} \right) = 0,
\end{align}
while ${V_0}\left( {{\xi _0},r} \right)$ can be expanded using the Fourier series by following coefficients: 
	\begin{align}\label{eq14}
{V_0}\left( {{\xi _0},r} \right) = \frac{1}{\pi }\int_{ - 1}^{ + 1} {V\left( {r + {\xi _0}\rho } \right)} \frac{{d\rho }}{{\sqrt {1 - {\rho ^2}} }},
\end{align}
Employing the Ehlotzky’s approximation \cite{ehlotzky1985scattering}:
	\begin{align}\label{eq15}
V\left( {r + {\xi _0}\rho } \right) + V\left( {r - {\xi _0}\rho } \right) \approx V\left( {r + {\xi _0}} \right) + V\left( {r - {\xi _0}} \right),
\end{align}
and hence, by evaluating the integral in (14), we obtain
	\begin{align}\label{eq16}
{V_0}\left( {{\xi _0},r} \right) = \frac{1}{2}\left[ {V\left( {r + {\xi _0}} \right) + V\left( {r - {\xi _0}} \right)} \right]
\end{align}
Equation \ref{eq16} is the approximate expression to model the laser field. Now we can write the Stationary-state Schrodinger equation by considering laser effect for potential $V\left( r \right)$ as follows:
	\begin{align}\label{eq17}
\left( {{\nabla ^2} + {k^2}} \right)\Psi _k^0\left( {\bf{r}} \right) = \frac{\mu }{{{\hbar ^2}}}\left[ {V\left( {r + {\xi _0}} \right) + V\left( {r - {\xi _0}} \right)} \right]\Psi _k^0\left( {\bf{r}} \right)
\end{align}
where $k = \sqrt {2\mu {E_{KH}}/{\hbar ^2}}$ is wave number. The equation \ref{eq17} represents motion for one spherically confined two particles exposed to linearly polarized intense laser field radiation. 

\section{Entanglement fidelity at the presence of a laser field}\label{3}
 In order to give a quantitative measurement for the entanglement, we introduce the entanglement fidelity (EF). First, we consider the scattering of a particle through the potential. The equation \ref{eq17} describes the stationary-state Schrodinger equation containing the potential which characterizes the quantum collision processes and laser effect. 
 
 The final state wave function $\Psi _k^0\left( {\bf{r}} \right)$ is represented by the partial wave expansion \cite{geltman2013topics,sitenko2013lectures} in the following form:
	\begin{align}\label{eq18}
\Psi _k^0\left( {\bf{r}} \right) = \sum\limits_{l = 0}^\infty  {{i^l}\left( {2l + 1} \right)} {D_l}(k){P_l}(\cos \theta ){R_l}(r)
\end{align}
where ${D_l}(k)$ is the expansion coefficient,$i$ is the pure imaginary number and${R_l}(r)$ is the solution of the radial wave equation:
	\begin{align}\label{eq19}
&\left\{ {\frac{1}{{{r^2}}}\frac{d}{{dr}}\left( {{r^2}\frac{d}{{dr}}} \right) - \frac{{l\left( {l + 1} \right)}}{{{r^2}}}} \right.\nonumber\\ 
&\left. { - \frac{\mu }{{{\hbar ^2}}}\left[ {V\left( {r + {\xi _0}} \right) + V\left( {r - {\xi _0}} \right)} \right] + {k^2}} \right\}{R_l}(r) = 0
\end{align}
 ${P_l}(\cos \theta )$ is the Legendre polynomial of order $l$, while $l$ is the angular momentum quantum number. For a spherically symmetric potential, it has been shown that the radial wave equation ${R_l}(r)$ and the expansion coefficient ${D_l}(k)$ are given by \cite{mishima2004entanglement,mogaddam2019entanglement}:
 \begin{widetext}	
	\begin{align}\label{eq20}
{D_l}(k) = {(2\pi )^{ - 3/2}}\left\{ {1 + \frac{{i\mu k}}{{{\hbar ^2}}}} \right.{\left. {\int_0^\infty  {dr{\mkern 1mu} {r^2}{j_l}(kr)\left[ {V\left( {r + {\xi _0}} \right) + V\left( {r - {\xi _0}} \right)} \right]{R_l}(r)} } \right\}^{ - 1}}
\end{align}

	\begin{align}\label{eq21}
\begin{array}{l}
{R_l}(r) = {j_l}(kr) + \frac{{\mu k}}{{{\hbar ^2}}}\left\{ {{n_l}\left( {kr} \right)\int_0^r {dr'{{r'}^2}} {j_l}(kr')\left[ {V\left( {r' + {\xi _0}} \right) + V\left( {r' - {\xi _0}} \right)} \right]{R_l}(r')} \right.\\
\,\,\,\,\,\,\,\,\,\,\,\,\,\,\,\,\,\,\,\,\,\,\,\,\,\,\,\,\,\,\,\,\,\,\,\,\,\,\,\,\,\,\,\,\,\,\,\,\,\,\,\,\,\,\,\,\,\,\,\left. { + {j_l}(kr)\int_r^\infty  {dr'{{r'}^2}} {n_l}(kr')\left[ {V\left( {r' + {\xi _0}} \right) + V\left( {r' - {\xi _0}} \right)} \right]} \right\}{R_l}(r')
\end{array}
\end{align}
\end{widetext}
Asymptotic form of the radial wave function can be achieved by the phase-shift ${\delta _l}$ such that ${R_l}\left( r \right) \propto {\left( {kr} \right)^{ - 1}}\sin \left( {kr - \pi l/2 + {\delta _l}} \right)$.
It has been shown that the collisional EF for the scattering process can be represented by ${f_k} \propto {\left| {{d^3}r\,\Psi _k^0\left( r \right)} \right|^2}$, which is the absolute square of the scattered wave function for a given interaction potential \cite{mishima2004entanglement}. In low energy collisions, the main contribution in the scattering process is related to the partial $s$ -wave scattering ($l = 0$). Therefore, the EF (i.e. ${f_k}$) can be calculated through the expansion coefficient ${D_l}(k)$ and the radial wave equation ${R_l}(k)$, as follows:
	\begin{align}\label{eq22}
{f_k} \propto \frac{{{{\left| {{D_0}\left( k \right)} \right|}^2}{{\left| {\int_0^\infty  {dr{\mkern 1mu} {r^2}{R_0}\left( k \right)} } \right|}^2}}}{{1 + {{\left| {\frac{{\mu k}}{{{\hbar ^2}}}\int_0^\infty  {dr{\mkern 1mu} {r^2}\left[ {V\left( {r + {\xi _0}} \right) + V\left( {r - {\xi _0}} \right)} \right]{R_0}\left( k \right)} } \right|}^2}}}
\end{align}
The above equation can explain most features of the collisional EF successfully. As a cross check, one can find that in the limit ${\xi _0} \to 0$ (when the influence of laser field vanishes) the Eq. (\ref{eq22}) reduces to the result (the equation (18) ) of the Mishima et al work \cite{mishima2004entanglement}.
\section{Effective potential of hot quantum plasmas}\label{4}
An analytical formulation for the effective interaction potential in hot quantum plasmas has been derived using the quantum approach including the influence of the effective plasma screening effects due to the collective plasma oscillations. The picture of a dressed Debye interaction between charged particles in hot quantum plasmas usually finds a complicated situation if we consider effective screening effects. Using the effective screening potential, the dressed electron-ion interaction potential ${V_{eff}}(r,\beta ,{\lambda _D})$ in hot quantum plasma can be written as \cite{,ramazanov2005runaway,jung2011plasmon,kvasnica1975bremsstrahlung}:
 \begin{widetext}	
	\begin{align}\label{eq23}
{V_{eff}}(r,\beta ,{\lambda _D}) =  - \frac{{Z{e^2}}}{r}\frac{1}{{4\sqrt {1 - {\beta ^2}} }}\left[ {\left( {4 - \beta } \right){e^{ - r/{L_1}\left( {\beta ,{\lambda _D}} \right)}} - 2\left( {1 - \sqrt {1 - {\beta ^2}} } \right){e^{ - r/{L_2}\left( {\beta ,{\lambda _D}} \right)}}} \right]{\mkern 1mu} {\mkern 1mu} 
	\end{align}
\end{widetext}
where $Z$ is the charge number of the ion,$\beta  = \hbar {\omega _p}/{k_\beta }T$ is the ratio of the plasmon energy $\hbar {\omega _P}$ to the thermal energy ${k_B}T$,${\omega _P}$ denotes the plasmon frequency, ${k_B}$ is the Boltzmann constant, $T$ stands for the plasma temperature, ${L_1}$ and ${L_2}$ are the effective screening lengths which are define as:
	\begin{align}\label{eq24}
{L_1} = \frac{{{\lambda _D}}}{{\sqrt 2 }}{\left( {1 + \sqrt {1 - {\beta ^2}} } \right)^{1/2}}{\mkern 1mu},\nonumber\\ 
{L_2} = \frac{{{\lambda _D}}}{{\sqrt 2 }}{\left( {1 - \sqrt {1 - {\beta ^2}} } \right)^{1/2}}{\mkern 1mu} 
\end{align}
while ${\lambda _D}$ is the standard Debye radius. This potential is expected to be valid when the plasmon energy is smaller than the thermal energy, i.e.,$0 \le \beta  < 1$. The effective potential ${V_{eff}}(r,\beta ,{\lambda _D})$ is a bright example for renomarlization process which can be applied to study the collective interactions in hot quantum plasmas. However, if the plasmon effects are neglected, the effective interaction potential ${V_{eff}}(r,\beta ,{\lambda _D})$ goes toward the classical Debye-Huckel potential:${V_{eff}} \to {V_{DH}} =  - \frac{{Z{e^2}}}{r}{\mkern 1mu} {\mkern 1mu} {e^{ - r/{\lambda _D}}}$, since ${L_1} \to {\lambda _D}$ and ${L_2} \to 0$ as $\beta  \to 0$. 
\section{EF of hot quantum plasmas by considering a laser field}\label{5}	
 The fidelity ratio is a powerful measure for investigating the plasmon and plasma screening effects on the entanglement fidelity for the elastic collisions in hot quantum plasmas. The fidelity ratio ${R_F} = f_k^{eff}/f_k^{Coul}$ is calculated by the ratio of the entanglement fidelity ${f_k}^{eff}$ for the elastic electron-ion collision (using the effective interaction potential ${V_{eff}}(r,\beta ,{\lambda _D})$ in hot quantum plasmas) to the entanglement fidelity ${f_k}^{Coul}$ for the elastic electron-ion collision using the pure Coulomb interaction ${V_C} =- \frac{{Z{e^2}}}{r}$. Thus, we have:
  \begin{widetext}
 	\begin{align}\label{eq25}
{R_F}\, = \frac{{1 + {{\left| { - \frac{{2\mu kZ{e^2}}}{{{\hbar ^2}}}\int_0^\infty  {dr{\mkern 1mu} r\frac{{\sin (kr)}}{{kr}}} } \right|}^2}}}{{1 + {{\left| {\frac{\mu }{{{\hbar ^2}}}\int_0^\infty  {dr{\mkern 1mu} r\left[ {{V_{eff}}(r + {\xi _0},\beta ,{\lambda _D}) + {V_{eff}}(r - {\xi _0},\beta ,{\lambda _D})} \right]\sin (kr)} } \right|}^2}}}.
 \end{align}
\end{widetext}

We arrive at the following relation for the fidelity ratio:
 	\begin{align}\label{eq26}
{R_F}\left( {\bar E,\beta ,{{\bar \lambda }_D},{\xi _0}} \right) = \frac{{1 + \frac{{4{Z^2}{\mu ^2}{e^4}}}{{{\hbar ^4}{k^2}}}}}{{1 + \frac{{4{Z^2}{\mu ^2}{e^4}}}{{{\hbar ^4}{k^2}}}{{\left| {{G_L}} \right|}^2}}}
\end{align}
in which
  \begin{widetext}
	\begin{align}\label{eq27}
\begin{array}{l}
{G_L}\, = \frac{{\left( {4 - \beta } \right)}}{{8\sqrt {1 - {\beta ^2}} }}\left[ {\left. {\frac{{2{k^2}\cosh \left( {{\xi _0}/{L_1}} \right)}}{{1/{L_1}^2 + {k^2}}} + k{\xi _0}{\mathop{\rm Im}\nolimits} \left\{ {{e^{ik{\xi _0}}}\,Ei\left( {1, - \left( {1/{L_1} - ik} \right){\xi _0}} \right) - {e^{ - ik{\xi _0}}}\,Ei\left( {1,\left( {1/{L_1} - ik} \right){\xi _0}} \right)} \right\}} \right]} \right.\\
\,\,\,\,\,\,\,\,\,\, - \frac{{k\left( {1 - \sqrt {1 - {\beta ^2}} } \right)}}{{4\sqrt {1 - {\beta ^2}} }}\left[ {\left. {\frac{{2{k^2}\cosh \left( {{\xi _0}/{L_2}} \right)}}{{1/{L_2}^2 + {k^2}}} + k{\xi _0}{\mathop{\rm Im}\nolimits} \left\{ {{e^{ik{\xi _0}}}\,Ei\left( {1, - \left( {1/{L_2} - ik} \right){\xi _0}} \right) - {e^{ - ik{\xi _0}}}\,Ei\left( {1,\left( {1/{L_2} - ik} \right){\xi _0}} \right)} \right\}} \right]} \right.\,
\end{array}
	\end{align}
\end{widetext}
We have used the following integration relation to calculate (\ref{eq26}):
	\begin{align}\label{eq28}
&\int_0^\infty  {\frac{r}{{r \pm \varepsilon }}\,{e^{ - \eta \left( {r \pm \varepsilon } \right)}}\sin \left( {kr} \right)dr}  = {\mathop{\rm Im}\nolimits} \left\{ {{e^{ \mp ik\varepsilon }}} \right. \times \nonumber\\
&\left. {\left[ {\frac{{\Gamma \left( {1, \pm \left( {\eta  - ik} \right)\varepsilon } \right)}}{{\eta  - ik}} \mp \varepsilon Ei\left( {1, \pm \left( {\eta  - ik} \right)\varepsilon } \right)} \right]} \right\}
\end{align}
	where
 	\begin{align}\label{eq29}
\Gamma \left( {a,x} \right) &= \int_x^\infty  {{e^{ - t}}{t^{a - 1}}dt};{\mathop{\rm Re}\nolimits} \left( a \right) > 0 \nonumber\\
Ei\left( {1,x} \right) &= \int_x^\infty  {\frac{{{e^{ - t}}}}{t}dt} 
\end{align}
It may be noted that $\Gamma \left( {a,x} \right)$ is called incomplete Gamma function and $Ei\left( {1,x} \right)$ is the Exponential integral [arfken].
In the limit ${\xi _0} \to 0$, the Eq. (\ref{eq26}) reduces to the derived equation by Jung \cite{jung2011plasmon} for a similar system but without considering the laser field.
\section{Numerical discussion}\label{6}	
\begin{figure}
	\begin{subfigure}{0.23\textwidth}\includegraphics[width=\textwidth]{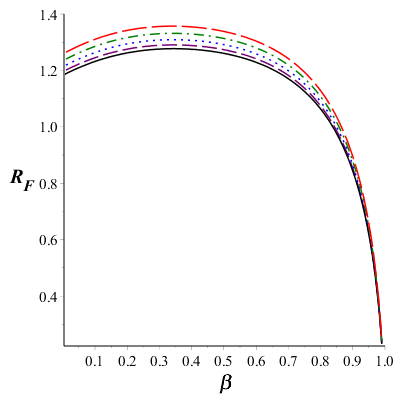}
		\caption{}
		\label{fig:1-1}
	\end{subfigure}
	\begin{subfigure}{0.23\textwidth}\includegraphics[width=\textwidth]{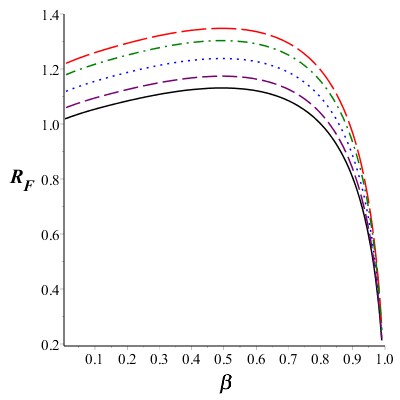}
		\caption{}
		\label{fig:1-2}
	\end{subfigure}
	\begin{subfigure}{0.23\textwidth}\includegraphics[width=\textwidth]{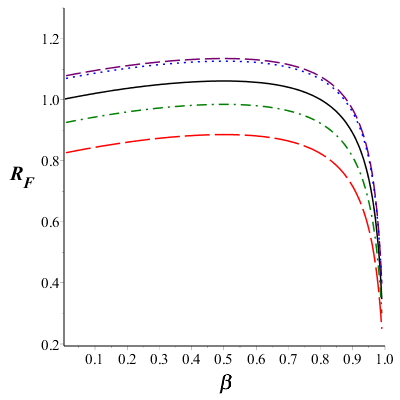}
		\caption{}
		\label{fig:1-3}
	\end{subfigure}
			\begin{subfigure}{0.23\textwidth}\includegraphics[width=\textwidth]{1c}
    	\caption{}
		\label{fig:1-4}
	\end{subfigure}
	\caption{The EFR ${R_F}$ including the plasmon, screening and laser effect as a function of the plasmon parameter $\beta $ when ${\bar \lambda _D} = 15$ and a) $\bar E = 0.05$, b) $\bar E = 0.5$, c) $\bar E = 5$ and d)$\bar E = 10$. In all panels we have chosen, solid line (${\xi _0} = 0$), dashed line (${\xi _0} = 0.005$), dotted line (${\xi _0} = 0.01$), dash-dot line (${\xi _0} = 0.015$), long dash (${\xi _0} = 0.02$).}
	\label{fig1}
\end{figure}

Analytical discussion on the behavior of the system is very difficult because of complicated functions in which physical parameters are involved. Therefore, we have setup several numerical calculations in order to understand different features of the system. All numerical calculations have been done using the following dimensionless variables:
	\begin{align}\label{eq30}
\bar E = {a_Z}^2{k^2},{\bar L_1} = \frac{{{L_1}}}{{{a_Z}}},{\bar L_2} = \frac{{{L_2}}}{{{a_Z}}},{\bar \lambda _D} = \frac{{{\lambda _D}}}{{{a_Z}}},{a_Z} = \frac{{{\hbar ^2}}}{{\mu Z{e^2}}}
\end{align}
The Amplitude of free electron oscillation in applied field, ${\xi _0} = e{E_0}/\mu {\omega ^2}$ is an important parameter. It is clear that ${\xi _0}$ is an increasing function of applied electric field and inversely proportional to the field angular frequency. In order to understand effects of ${\xi _0}$, we have plotted ${R_F}$ for several values of scaled collision energy $\bar E$. Figure \ref{fig1} presents fidelity ratio ${R_F}$ as functions of $\beta $ (ratio of plasmon energy to the thermal energy) for different values of ${\xi _0}$ and $\bar E$. Figures \ref{fig1} show that, general behavior of ${R_F}$ is the same for all values of scaled collision energy ($\bar E$) and the free electron oscillation amplitude (${\xi _0}$). Fidelity rises to a maximum value and rapidly fall as $\beta $ increases. Maximum value of fidelity of the system increases when the scaled collision energy increases. This means that the entanglement between particles reduces, as the collision energy is increased. Variation of ${R_F}$ respect to ${\xi _0}$ is more noticeable with greater values of $\bar E$ as one can find from figures 1.  

\begin{figure}
	\begin{subfigure}{0.23\textwidth}\includegraphics[width=\textwidth]{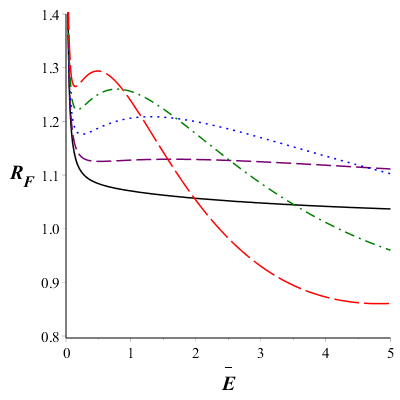}
		\caption{}
		\label{fig:2-1}
	\end{subfigure}
	\begin{subfigure}{0.23\textwidth}\includegraphics[width=\textwidth]{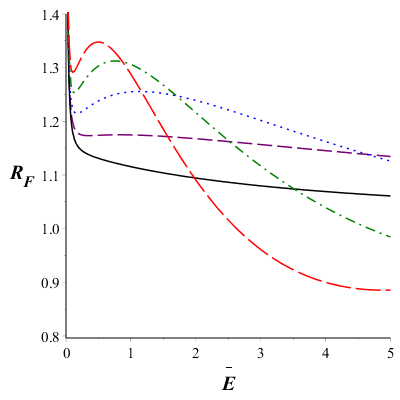}
		\caption{}
		\label{fig:2-2}
	\end{subfigure}
	\begin{subfigure}{0.23\textwidth}\includegraphics[width=\textwidth]{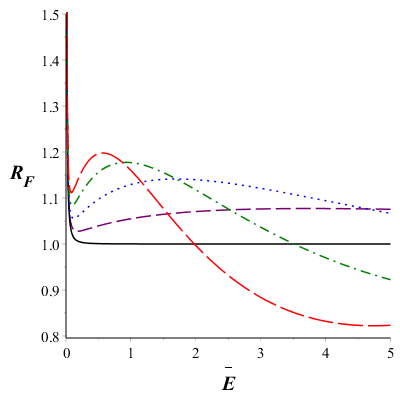}
		\caption{}
		\label{fig:2-3}
	\end{subfigure}
	\caption{The EFR ${R_F}$ including the plasmon, screening and laser effect as a function of the scaled collision energy $\bar E$ when ${\bar \lambda _D} = 15$ and plasmon parameter a) $\beta  = 0.2$,b) $\beta  = 0.5$, c) $\beta  = 0.8$. In all panels we have chosen, solid line (${\xi _0} = 0$), dashed line (${\xi _0} = 0.005$), dotted line (${\xi _0} = 0.01$), dash-dot line (${\xi _0} = 0.015$), long dash (${\xi _0} = 0.02$).}
	\label{fig2}
\end{figure}

For finding better view about the behavior of ${R_F}$, we have plotted fidelity as functions of the scaled collision energy $\bar E$ for different values of $\beta $ and ${\xi _0}$ in figure \ref{fig2}.

\begin{figure}
	\begin{subfigure}{0.23\textwidth}\includegraphics[width=\textwidth]{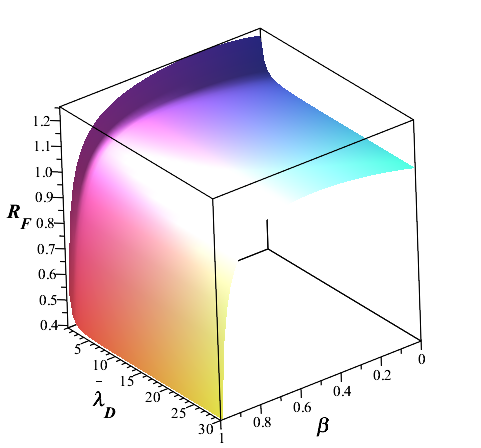}
		\caption{}
		\label{fig:3-1}
	\end{subfigure}
	\begin{subfigure}{0.23\textwidth}\includegraphics[width=\textwidth]{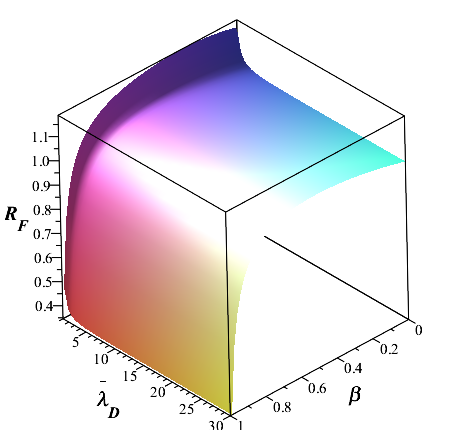}
		\caption{}
		\label{fig:3-2}
	\end{subfigure}
	\caption{The surface plot of the EFR ${R_F}$ as functions of the scaled Debye length ${\bar \lambda _D}$ and plasmon parameter $\beta $ when $\bar E = 5$ and a) ${\xi _0} = 0.01$ or b) ${\xi _0} = 0$.}
	\label{fig3}
\end{figure}

Figures \ref{fig2} indicate that Fidelity in the system destroys as collisional energy $\bar E$   increases. On the other hand, increasing the free electron oscillation amplitude ${\xi _0}$ also another reason of fidelity lost. In other words, both collision energy and free electron oscillation energy leading to decreasing the entanglement in the system. One can find from figures \ref{fig2} that, we can optimize the fidelity ratio ${R_F}$ by taking suitable values for ${\xi _0}$ and $\bar E$ parameters, which is an interesting result.
Another important parameter is the scaled Debye length ${\bar \lambda _D}$. Effects of this parameter on the entanglement of the system can be explained using the figure \ref{fig3}.

Figures \ref{fig3} show that fidelity ratio, slightly decreases when  increases. In distances smaller than the Debye length, long range effects are screened. Therefore, it is expected that entanglement effects due to long range interactions decrease as the Debye length increases. On the other hand, the distance between entangled atoms are very smaller than the Debye length, thus screening effect in the entanglement between plasma particles is small as we learn from the figure \ref{fig3}.

\begin{figure}
 	\begin{subfigure}{0.23\textwidth}\includegraphics[width=\textwidth]{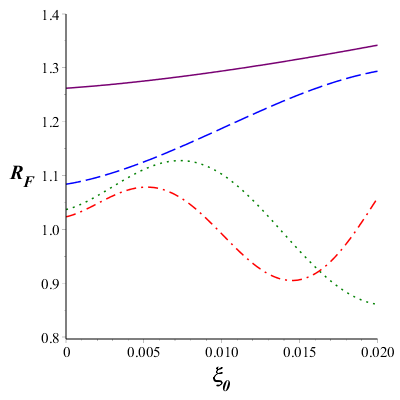}
 		\caption{}
 		\label{fig:4-1}
 	\end{subfigure}
 	\begin{subfigure}{0.23\textwidth}\includegraphics[width=\textwidth]{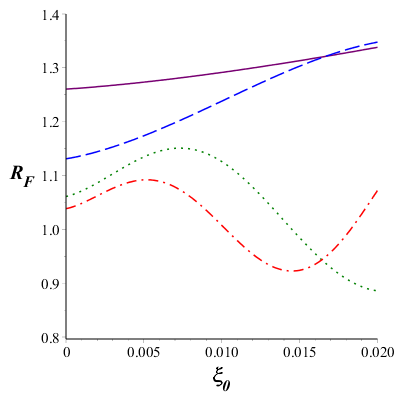}
 		\caption{}
 		\label{fig:4-2}
 	\end{subfigure}
  	\begin{subfigure}{0.23\textwidth}\includegraphics[width=\textwidth]{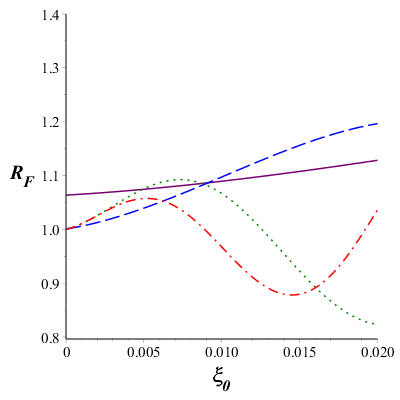}
 	\caption{}
 	\label{fig:4-3}
 \end{subfigure}
 	\caption{The EFR ${R_F}$ including the plasmon, screening and laser effect as a function of the ${\xi _0}$, for $\bar E = 0.05$ (Solid line), $\bar E = 0.5$ (dashed line), $\bar E = 5$ (dotted line), $\bar E = 10$ (dash-dot line). Plasmon parameter is   a) $\beta  = 0.2$,b) $\beta  = 0.5$, c) $\beta  = 0.8$ while ${\bar \lambda _D} = 15$ in all case.}
 	\label{fig4}
 \end{figure}

The figures \ref{fig4} provide better view to understand the effect of the amplitude of free electron oscillation (${\xi _0}$), the collisional energy $\bar E$ and $\beta$ on the behavior of entanglement fidelity ${R_F}$. Figures \ref{fig4} show that the ${R_F}$ generally is a periodic function of ${\xi _0}$ which its amplitude and frequency are functions of $\bar E$ and $\beta$. The oscillation frequency and its amplitude increase as collision energy increases. For a fixed value of ${\xi _0}$, variation of ${R_F}$ decreases as maximum value of $\beta$ increases.   

\section{Conclusions}\label{7}	
   In this work, we have studied effects of electric field due to the applied laser beam on the effective potential, describing bi-partite systems. 
We have used the Kramers-Henneberger unitary transformation, in the quantized field formalism while the squared vector potential (which appears in the equation of motion) is eliminated. The resultant equation is expressed in the Kramers-Henneberger frame and the corresponding wave function are expanded using the Fourier series by considering the Ehlotzky’s approximation. 
The stationary-state Schrodinger equation containing laser effect has been calculated which represents motion for one spherically confined two particles exposed to linearly polarized intense laser field radiation in center of mass coordinate. The partial wave method is employed to obtain the entanglement fidelity containing laser beam effect as well. Our calculations indicate that high intense laser fields change the behavior of the entanglement fidelity extensively.
We have studied some simple examples of the effective potential energy function to understand free electron oscillation amplitude and its relation with plasmon and quantum screening effect on the entanglement fidelity. Our results obtain that the laser effect, plasmon coupling and screening effects play a significant role in the entanglement fidelity for the elastic electron-ion collisions in hot quantum plasmas. 
These results should provide useful information on the laser- plasma interaction, quantum screening effect and the quantum information. 
\section{Acknowledgment}\label{8}
This work was partially supported by the Ferdowsi University of Mashhad under Grant No. 3/43953.
\bibliography{refs}
\end{document}